\documentclass{jnmp}

\usepackage{amsmath}

\JNMPnumberwithin{equation}{section}

\begin{document}

\renewcommand{\evenhead}{H Leblond}
\renewcommand{\oddhead}{Higher Order Terms in Multiscale Expansions:
A Linearized KdV  Hierarchy}

\thispagestyle{empty}

\FirstPageHead{9}{3}{2002}{\pageref{Leblond-firstpage}--\pageref{Leblond-lastpage}}{Article}

\copyrightnote{2002}{H  Leblond}

\Name{Higher Order Terms in Multiscale Expansions: \\
A Linearized KdV  Hierarchy}\label{Leblond-firstpage}

\Author{Herv\'e LEBLOND}

 \Address{Laboratoire POMA, UMR CNRS 6136, Universit\'e d'Angers,\\
 2 ${\it B^d}$ Lavoisier 49045 ANGERS Cedex 01, France\\
E-mail: herve.leblond@univ-angers.fr}

\Date{Received November 22, 2001; Revised February 1, 2002; Accepted
February 27, 2002}

\begin{abstract}
\noindent We consider a
wide class of model  equations,
able to describe wave propagation in dispersive nonlinear media.
The Korteweg-de Vries
(KdV) equation is derived in this general frame under some conditions,
the physical meanings of which are clarified.  It
is obtained as usual at leading order in some  multiscale expansion.
The higher order terms in this expansion
are studied making use  of a multi-time formalism
and imposing the condition that the main term  satisfies the whole KdV hierarchy.
The evolution of the higher order terms with respect to the
higher order time variables can be described through
the introduction of a linearized KdV hierarchy. This allows one
 to give an expression of the higher order time derivatives
that appear in the right hand  member of the perturbative expansion equations,
 to show that  overall the higher order terms  do not produce
 any secularity and to prove that the formal expansion contains only bounded terms.
\end{abstract}

\section{Introduction}

Soliton theory is based on the resolution of the so-called ``completely  integrable
equations'' by means of the inverse scattering transform (IST) method.
When they describe  particular physical
situations, these equations arise as asymptotics of well-established models,
and are derived from the latter by
 means of a multiscale expansion, or some equivalent formalism.
The integrable nonlinear evolution equations
appear as the leading order approximation in this perturbative approach.
Corrections to this first approximation are often to be considered.
At a given propagation time (or distance), more accuracy can
obviously be obtained
by retaining more than the main term only in the power series.  Such
additional terms are the ``higher order terms''.
On the other hand, for a very long propagation time, the physical solution goes away from
the theoretical soliton because the nonlinear evolution equation gives only
a first order approximation of this
evolution. Corrections to the equation must be taken into account.  The partial
differential equations giving such corrections are what we call the ``higher order'' ones.

Consideration of such corrections
 have given rise to perturbation theories for solitons~\cite{Karpman}
 that have found
 applications,  e.g.\ in the frame of the physics of optical solitons in fibres, which are
related to  optical telecommunications~\cite{kodama}.
The effect of the higher order equations appears
in these theories notably through the renormalization of the soliton velo\-city.
From the purely theoretical point of view, it has been recently shown that the
solvability
of the higher order equations can in some sense be related to the problem
of the complete integrability of the basic system~\cite{kodamaobs,manna2}.
Furthermore, a physical interpretation of the equations of the
Korteweg-de Vries  (KdV) hierarchy has
been found
by Kraenkel, Manna and Pereira in the framework of the theory
of water waves~\cite{manna4,manna1,manna3}.

 The Korteweg-de Vries (KdV) equation, which is known
to describe the propagation of  long waves in shallow water~\cite{kdvpap},
 is the first equation that was solved by the IST method~\cite{kdvint,kdvint2}.
More physically speaking,  the
first  observed soliton was a long wave in shallow water.
The KdV equation appears as the evolution equation satisfied by the dominant
term of the quantity describing the wave  in some multiscale expansion.
We investigate in this paper the higher orders
terms of such an expansion.
Our results apply {\it a priori}
to the Maxwell--Landau model that describes electromagnetic wave
 propagation in ferromagnetic media,
but, because they do not involve the explicit computations
 particular to  this situation (written down elsewhere~\cite{kdv2fer}),
 they are presented
in a general abstract frame.
The results are thus expected to  apply to other  physical situations.
  The  conditions under which the
KdV model can be derived in this general frame are discussed below
from both the physical
and the mathematical points of view.
Regarding the higher order equations
such conditions
cannot be written down explicitly in a completely satisfactory way
 from the mathematical
point of view, but their physical meaning is clarified. Further
 it has been proved~\cite{kdv2fer} that these conditions  are satisfied
by the Maxwell--Landau model which describes wave propagation in ferromagnets.

We use, as did Kraenkel {\it et al}, a multiple time formalism.
These authors have shown that the evolution of the dominant term
relative to the higher order time variables is governed by the
KdV hierarchy. Kodama and Taniuti showed that the evolution of the
higher order terms relative to the first time variable is governed
by the linearized KdV equation, but
the way in which the higher order terms depend on the higher order time
 variables was not yet clarified and
it was not  proven that the higher
order terms do not produce secular terms,
while the coherence of the multiscale expansion necessitates the elimination
of any secularities. We find that the  evolution of these terms can
be described by means of some linearized KdV
hierarchy. This is the main result of the paper
and enables us to prove that no unbounded term does appear in the
expansion.

The  paper is organized as follows.  In Section 2 we describe
 the frame  of the multiscale expansion, derive the KdV equation and
discuss the physical hypotheses that allow this derivation.
 In Section~3
the problem of secularities is presented
and the introduction of the KdV hierarchy allows us
both to remove the secularity producing terms due to the main
order and to determine the evolution of the latter relative to the
higher order variables.
Section~4 is devoted to the higher order time evolution of the
higher order terms and to the linearized KdV hierarchy.
Section~5 contains a conclusion.
The linear part of the right hand member of the equations of the perturbative expansion
plays an important role in the treatment of the secularity producing terms.
Therefore the corresponding coefficients are computed
 explicitly in an appendix.

\newpage

\section{A multiscale expansion}
\subsection{The model and the scaling}

We consider some set of partial differential equations
that can be written as
\begin{equation}\left(\partial_t + A\partial_x +E\right)u=B(u,u),
\label{3.1}\end{equation}
where the function $u$ of the variables $ x$ and $t$ is valued in
${\mathbb R}^p$,
$A $ and $E$ are some $p\times p$
 matrices, and $B:{\mathbb R}^p\times{\mathbb R}^p\to{\mathbb R}^p$
is bilinear. This system can describe the propagation of electromagnetic waves
 in
a ferromagnetic medium, according to the Maxwell and Landau equations~\cite{kdv2fer}.
We consider the abstract system~(\ref{3.1}) rather than this latter particular situation
for sake of simplicity.
Indeed
the study of a specific case would imply explicit computation of all
the quantities involved by the multiscale expansion and
 increase consi\-de\-rably the size of the expressions,
while the matter of the present paper does not depend upon the particular
form of these quantities. It is in fact  rather general, according to the fact that
  KdV solitons arise in many other physical frames.
Considering an abstract frame thus avoids  much computational detail,
but in fact is not so easy because the derivation of the KdV model
is not always achievable. It
involves several assumptions, some of which are rather strong.
These assumptions are introduced
 at the points where they  have to be used.
Then in subsection~\ref{II.3} we give their physical interpretation.
In any case the required assumptions
are satisfied by the Maxwell--Landau model, describing waves in
ferromagnets. The proof of this is given in~\cite{kdv2fer}.
Thus the results of the paper are valid and completely justified  for
this latter model. The use of the abstract system~(\ref{3.1})
presents the two following advantages.  Firstly
it simplifies the algebra because it avoids considering many details
peculiar to the physics of ferromagnets. Secondly it allows a~discussion
of the physical conditions required for the occurrence of  KdV solitons, despite
some feature concerning the higher order equations
that cannot be completely solved from the mathematical point of view.

Regarding the system (\ref{3.1}) the following assumptions are usually made:
$A$  is assumed either to be
 completely hyperbolic or   symmetric and $E$ is assumed to be skewsymmetric.
The two latter hypotheses are closely related to the conservative character of
the system and are
  satisfied by the Maxwell--Landau system under the scalar product
\[
\left(\left(\vec E, \vec H, \vec M\right)\bigl|\left(\vec E, \vec H, \vec M\right)\right)
=\vec E^2+ \vec H^2+\alpha \vec M^2
\]
 (with the notations of \cite{kdv2fer}).
Note that this scalar product depends upon the zero-order term.
Only weaker assumptions are necessary for the formal derivation of the KdV equation,
as we will see below.
However,  some essential symmetry property of the higher order equations
is satisfied only when the system is conservative.
 This is ensured,
regarding the linear part of the system,
by the symmetry and skewsymmetry  hypotheses above.
The corresponding assumption on the nonlinear part cannot be expressed in a simple
way.

The variable $u$ is expanded in a power series of some small parameter $\varepsilon$ as
\begin{equation}
u=\varepsilon^2 u_2+\varepsilon^3 u_3+\cdots,
\label{12}\end{equation}
and  the slow variables
\begin{equation} \xi=\varepsilon(x- V t),\qquad
\tau=\varepsilon^3t\label{13}\end{equation}
 are introduced.
 $ V$ is a speed to be determined.
This multiscale expansion is well-known. It yields the Korteweg-de
Vries (KdV) equation at leading order, $\varepsilon^2$, in many physical
cases. Starting the expansion at order $\varepsilon^2$ corresponds
physically to specifying a low value for the order of magnitude of
the amplitude. Setting a term of order $\varepsilon$ in expansion~(\ref{12})
is {\it a priori} possible, but it is easily checked
that it does not yield any evolution equation regarding the chosen
time scales. Therefore we can omit it. It is seen below that, as
is described in~\cite{kodamah,manna3},
  secularities, i.e.\  linear growth
of the solutions with time, appear  in the
higher order terms.
These
  secularities must be removed and this is achieved by imposing
to the leading term some particular dependency
with regard  to higher order time variables (the equations of the KdV hierarchy).
Therefore we introduce the
  variables $\tau_2, \tau_3,\ldots$  which are defined by
\begin{equation}
\tau_j=\varepsilon^{2j+1}t\qquad (j\geqslant1).\end{equation}
With this notation $\tau=\tau_1$.

\subsection{The Korteweg-de Vries equation}
At order $\varepsilon^2$ equation (\ref{3.1}) yields
\begin{equation} E u_2=0.\label{e5}\end{equation}
 Thus $u_2$
belongs to the kernel, ${\rm ker} (E)$, of $E$. We assume  that the
range, ${\rm Rg} (E)$, of $E$ is in direct sum with its kernel
 (hypothesis 1). This is satisfied if $E$ is skewsymmetric.
We also  denote by $\Pi_0$ and $Q_0$ respectively the projectors onto ${\rm ker} (E)$
parallel to  ${\rm Rg} (E)$  and onto  ${\rm Rg} (E)$  parallel to  ${\rm ker} (E)$.
Equation (\ref{e5}) can be written as
\begin{equation} u_2=\Pi_0 u_2.\label{e5.1}\end{equation}

At the following order, $\varepsilon^3$, we get
the equation
\begin{equation} Eu_3+(A-V)\partial_\xi u_2=0.\label{e7.3}\end{equation}
The projection of equation (\ref{e7.3}) on ${\rm ker}(E)$ is
\begin{equation}\Pi_0(A-V)\partial_\xi u_2=0.\label{e7.4}\end{equation}
Because of (\ref{e5.1}),
 equation (\ref{e7.4}) has a nonzero solution if $V$ is an eigenvalue of $\Pi_0A\Pi_0$.
We assume that $V$ is a simple  nonzero eigenvalue of $\Pi_0A\Pi_0$,
 and call $a_0$ an eigenvector,
$\Pi_1$ and $Q_1$ the associated projectors:
$\Pi_1$ the projector on $a_0\mathbb R$,
$Q_1$ the projector on ${\rm Rg}\left(\Pi_0(A-V)\Pi_0\right)$, so that
$\Pi_1+Q_1=\Pi_0$.
This definition assumes that ${\mathbb R}^p$ is the direct sum of the  range and the kernel of
$\Pi_0(A-V)\Pi_0$. This implies that not only the eigenspace but the
characteristic space of the operator $\Pi_0A\Pi_0$ relative to the eigenvalue $V$
has dimension~1 (hypothesis~2).
The assumption of the complete hyperbolicity of $A$ ensures that this hypothesis is
satisfied. When $A$ is assumed to be symmetric, the dimensions of characteristic and
eigenspace are the same.

According to equation (\ref{e7.4})
$\partial_\xi u_2\in{\rm ker}\left(\Pi_0(A-V)\Pi_0\right)$.
Thus $u_2=a_0\varphi_2$, where the function $\varphi_2$ has to be determined.
The $Q_0$-projection of equation (\ref{e7.3}) is
\begin{equation} Q_0u_3=a_1\partial_\xi\varphi_2.\end{equation}
The vector coefficient $a_1$ is defined by $a_1=-E^{-1}(A-V)a_0$,
where   $E^{-1}$
is a  partial inverse of $E$.
$E^{-1}$ is precisely defined as follows.  We
call $\hat E$ the restriction and corestriction
of $Q_0 EQ_0$ to
${\rm Rg} (E)$: $\hat E$ is invertible. We denote by $\hat 0$
the zero operator defined on ${\rm ker}(E)$.
Then $E^{-1}$ is the direct sum of $\hat 0$ and $\hat E^{-1}$.
In other words, the part of the matrices of $E$ and $E^{-1}$ corresponding to
${\rm Rg}(E)$ are inverses one of each other and the matrix of $E^{-1}$ is completed
by zeros to make a $p\times p$ matrix.

The equation of order $\varepsilon^4$
is
\begin{equation} Eu_4+(A-V)\partial_\xi u_3=B(u_2,u_2).\label{e8}\end{equation}
It is divided into 3 parts by using the projectors $Q_0$, $Q_1$,
$\Pi_1$. The   $\Pi_1$-projection is a~compatibility condition for
$\varphi_2$. Using $\Pi_1(A-V)\Pi_0=0$, and splitting $u_3$ into
$u_3=\Pi_0 u_3+Q_0u_3$, we find that the projection reduces to
\begin{equation} q\partial_\xi^2\varphi_2=r\varphi_2^2,\label{e9}\end{equation}
with $q=\Pi_1(A-V)a_1$ and $r=\Pi_1B(a_0,a_0)$.
If $(q,r)\neq(0,0)$, equation (\ref{e9}) has no bounded nonzero solution.
The following conditions must thus be satisfied (hypotheses 3 and 4),
{\it viz}
\begin{equation}\Pi_1 B(a_0,a_0)=0,\qquad\mbox{and} \qquad
\Pi_1 (A-V)a_1=0\label{con}.\end{equation}
The $Q_1$- and then the $Q_0$-projection of  equation (\ref{e8})  yield expressions for
$Q_1u_3$ and~$Q_0u_4$, respectively, as
\begin{gather} Q_1 u_3=a'_1\partial_\xi\varphi_2+\alpha_1\int^\xi{\varphi_2}^2,\label{e11.1}\\
 Q_0 u_4=a_1\partial_\xi\varphi_3+a_2\partial_\xi^2\varphi_2+
\alpha_2{\varphi_2}^2\label{e11.2}.\end{gather}
Denoting by
 $(A-V)^{-1}$
 a partial inverse of $\Pi_0(A-V)\Pi_0$ defined in an way analogous to~$E^{-1}$,
we define the vector coefficients by
\begin{gather} a'_1=-(A-V)^{-1}Q_1(A-V)a_1,\label{e11.3}\\
 \alpha_1=(A-V)^{-1}Q_1B(a_0,a_0),\label{e11.4}\\
  a_2=-E^{-1}Q_0(A-V)\left(a_1+a'_1\right),\label{e11.5}\\
   \alpha_2=E^{-1}Q_0\left(B(a_0,a_0)-(A-V)\alpha_1\right).\label{e11.6}\end{gather}

With the use of all previous results the compatibility condition
($\Pi_1$-projection) of the equation of
order $\varepsilon^5$ yields the following evolution equation for $\varphi_2$:
\begin{equation} \partial_\tau\varphi_2+\beta\varphi_2\partial_\xi\varphi_2
+\gamma\partial_\xi^3\varphi_2=\delta\varphi_2\int^\xi{\varphi_2}^2.
\label{e11.7}\end{equation}
The scalar coefficients $\beta$, $\gamma$, $\delta$ are defined by
\begin{gather} a_0\beta=2\Pi_1(A-V)\alpha_2-2\Pi_1 B(a_0,a_1+a'_1), \label{e11.8}\\
 a_0\gamma=\Pi_1(A-V)a_2,\label{e11.9}\\
  a_0\delta=2\Pi_1 B(a_0,\alpha_1). \label{e11.10}\end{gather}
If the additional condition (hypothesis 5)
\begin{equation}\Pi_1 B(a_0,\alpha_1) =0\label{e11.11}
\end{equation} is satisfied, equation (\ref{e11.7})
reduces to  the
KdV equation
\begin{equation} \partial_\tau\varphi_2+\beta\varphi_2\partial_\xi\varphi_2
+\gamma\partial_\xi^3\varphi_2=0.\label{kdv}\end{equation}

\subsection{Physical meaning of the hypotheses}
\label{II.3}
The above formal derivation of the KdV equation necessitates 5 hypotheses that have
been written down explicitly. Two of them, hypotheses 3 and 5, involve the nonlinearity, the
three other  the derivatives only.
A physical interpretation of the latter can be found in the dispersion relation
of the linearized system
\begin{equation}\left(\partial_t + A\partial_x +E\right)u=0.\label{3.1bis}\end{equation}
The pulsation and polarization vector corresponding
to  a given wave vector, $k$, are denoted by $\omega(k)$ and $u(k)$ respectively.
They satisfy
 \begin{equation}\left(-i\omega(k) + Aik +E\right)u(k)=0.\label{3.1ter}\end{equation}
The long wave approximation corresponds to
$k=0$ and $\omega(0)=0$. Then equation (\ref{3.1ter}) reduces to
$E\,u(0)=0$. It is equation (\ref{e5}).
Taking the derivative of equation (\ref{3.1ter})
with regard to $k$ and then
setting $k=0$ we obtain
\begin{equation} i[A-\omega'(0)]u(0)+E\,u'(0)=0.\label{z1}\end{equation}
The prime denotes the derivative with regard to $k$.
Together with the interpretation of the long wave approximation
as a limit of oscillating waves for $k$ close to~0,
the analogy between equation (\ref{z1}) and equation (\ref{e7.3})
allows the identification between $u(0)$ and~$a_0$ on one hand, and
between $\omega'(0)$ and $V$ on the other.
$V$~appears as a long-wave limit of the group velocity.
Hypothesis~2 is thus that a unique  polarization can propagate with
this velocity. If this assumption is not satisfied, interactions between
the various waves with same velocity must be taken into account.

Equation (\ref{z1}) yields also
 \begin{equation} Q_0u'(0)=-E^{-1} i[A-V]a_0.\label{z2}\end{equation}
Thus $Q_0u'(0)=ia_1$, with the  definition above of $a_1$.
Taking once  again the derivative of equation (\ref{3.1ter})
 with regard to $k$ we obtain, for $k=0$,
\begin{equation} -i\omega''(0)a_0+2i[A-V]u'(0)+E\,u''(0)=0.\label{z3}\end{equation}
Taking the $\Pi_1$-projection of equation (\ref{z3}),
decomposing $u'(0)$ into $u'(0)=ia_1+\Pi_0u'(0)$, and taking into account the
relation $\Pi_1(A-V)\Pi_0=0$, we get
\begin{equation} \omega''(0)a_0=2i\Pi_1[A-V]a_1.\label{z4}\end{equation}
Hypothesis 5, in (\ref{con}), is thus \[
\frac{d^2\omega}{dk^2}(0)=0.\] The following feature can be
observed: $\omega'''(0)$ can be computed by the same  method,
using the second derivative of equation (\ref{3.1ter}). The
expression obtained gives an interpretation of the coefficient
$\gamma$ of the third derivative in the KdV equation,
(\ref{e11.7}), as
\begin{equation}\gamma=\frac{-1}6\frac{d^3\omega}{dk^3}(0).\end{equation}

 Hypothesis 1 appears mainly as a technical assumption without special physical
meaning needed to solve equation (\ref{z2}).
If $E$ is skewsymmetric, this hypothesis is satisfied.
Assume that $E$ has a nonzero symmetrical part $E_s$.
Taking the scalar product of equation (\ref{3.1ter}) by $u(k)$,  then
comparing the result to its complex conjugate and making use of the
symmetry assumptions ($A$ being symmetric), we obtain  the following
expression for the imaginary part  $\omega_i(k)$ of $\omega(k)$:
$ \omega_i(k)={-(u|E_su)}/{(u|u)}$.
Thus the system cannot be conservative if $E$ is not skew-symmetric, and
hypothesis 1 is always satisfied by conservative systems.

The two hypotheses  3 and 5 concerning the nonlinearity obviously cannot be understood
through the study of the linearized equation (\ref{3.1bis}).
Hypothesis 3 can be written as
\begin{equation} \Pi_1B(\Pi_1,\Pi_1)=0,\label{z8}\end{equation}
while the  assumption
\begin{equation} \Pi_1B(\Pi_1,Q_1)=0\label{z9}\end{equation} implies hypothesis 5.
Conditions (\ref{z8}) and (\ref{z9}) are particular expressions
of a very general condition called the ``transparency'' condition
in the rigorous mathematical theory of multiscale expansions by Joly, M\'etivier and Rausch
\cite{jmr}. Condition (\ref{z8}) excludes quadratic self-interaction
for the chosen propagation mode, while condition (\ref{z9}) excludes
interaction at the same order for different polarizations.
If condition (\ref{z8}), that is hypothe\-sis~3, is not satisfied, the
nonlinear term appears sooner in the expansion.
The nonlinear evolution of the wave will be described by
 a nonlinear evolution equation other than the KdV equation
and for other space, amplitude and time scales.
If condition (\ref{z9}), or more precisely hypothesis~5, is not satisfied,
the above computation is valid, but the evolution equation is equation (\ref{e11.7}) instead of KdV.
It is an integro-differential equation involving a cubic nonlinear term.

\subsection{The linearized KdV equation}
Regarding the higher order terms,
 the evolution equation relative to the first order time $\tau=\tau_1$
is found in the following way.
Equation (\ref{3.1}), written for some given order $\varepsilon^n$, yields
a compatibility condition, its $\Pi_1$-projection
\begin{equation} \Pi_1(A-V)\partial_\xi Q_0 u_{n-1}+\sum_{j\geqslant 1}
\partial_{\tau_j}\Pi_1 u_{n-2j-1}=\sum_{p=2}^{n-2}
\Pi_1 B(u_p,u_{n-p}),\label{pieq}\end{equation}
and  recurrence formulas for the determined parts of $u_n$,
its $Q_0$-projection
\begin{equation} Q_0u_n=E^{-1}Q_0\left[-(A-V)\partial_\xi u_{n-1}-\sum_{j\geqslant 1}
\partial_{\tau_j} u_{n-2j-1}+\sum_{p=2}^{n-2}
B(u_p,u_{n-p})\right],\label{qeq}\end{equation}
and its $Q_1$-projection
\begin{gather} Q_1 u_n=(A-V)^{-1}Q_1\nonumber\\
\phantom{Q_1 u_n=}{} \times \left[
-(A-V)Q_0u_n-\sum_{j\geqslant1}\int^\xi\partial_{\tau_j}u_{n-2j}+
\sum_{p=2}^{n-1}\int^\xi B(u_p,u_{n+1-p})\right].\label{e15.1}\end{gather}
The component of $u_n$ belonging to ${\rm Rg}(\Pi_1)$ is
proportional to some unknown real function~$\varphi_n$, and $u_n$
is
\begin{equation} u_n=a_0\varphi_n+Q_0u_n+Q_1u_n.\end{equation}
We call ${\cal S}_p$ the set of expressions involving
$u_2, u_3,\ldots,u_p$, but not $u_{p+1}$ and subsequent terms.
The same notation ${\cal S}_p$ will hold below for any function belonging to this set.
Due to
equations (\ref{qeq}) and (\ref{e15.1})  $u_n$ can be written as
\begin{equation} u_n=a_0\varphi_n+{\cal S}_{n-1}.\label{un1}\end{equation}
The functions $\varphi_n$ are the unknowns of the problem.
Note that,  with these notations,  ${\cal S}_{p}$~is the
set of expressions involving $\varphi_2, \varphi_3,\ldots,\varphi_p$
and their derivatives.
Applying two times
the recurrence formulas (\ref{qeq}) and  (\ref{e15.1}) onto the expression (\ref{un1})
of $u_n$ we obtain
\begin{equation} u_n=a_0\varphi_n+Q_0u_n+Q_1u_n,\label{e18.1}\end{equation}
with
\begin{gather} Q_0u_n=
a_1\partial_\xi\varphi_{n-1}
+a_2\partial_\xi^2\varphi_{n-2}+2\alpha_2\varphi_2\varphi_{n-2}+
{\cal S}_{n-3},\label{e18.2}\\
 Q_1u_n=
a'_1\partial_\xi\varphi_{n-1}+
2\alpha_1\int^\xi\varphi_2\varphi_{n-1}+
{\cal S}_{n-2}.\label{e18.3}\end{gather}
Equations (\ref{e18.1})--(\ref{e18.3}) are correct if $n-2>2$, i.e.\  $n\geqslant5$. For $n=4$,
the coefficient 2 before $\alpha_{2}$ vanishes, as seen above. The dependency of
$Q_1u_n$ with regard to $\varphi_{n-2}$ is not needed in the computation
of the higher order equations below.
This is fortunate because of several integral nonlinear terms arising in it that
would greatly hinder the computation.
When use is made of the expressions (\ref{e18.1}) to (\ref{e18.3}) of  $u_n$
in the compatibility condition~(\ref{pieq}),
it gives an evolution equation for $\varphi_{n-3}$.
Indeed, $u_n$ does not appear in the equation. The
dependency with regard to $\varphi_{n-1}$, $\varphi_{n-2}$ and $\varphi_{n-3}$
is explicitly written. Because of the
condition $\Pi_1 (A-V)a=0$, which determines  the
velocity~$V$, $\varphi_{n-1}$ vanishes from the equation.
Because of the
relations~(\ref{con}) and~(\ref{e11.11}) that are satisfied under the present hypotheses,
 $\varphi_{n-2}$
and a term
$\delta\varphi_2\left(\int^\xi\varphi_2\varphi_{n-3}+\varphi_{n-3}\int^\xi{\varphi_2}^2\right)$
also vanish and the   equation obtained is, with $n-3=l$:
\begin{equation} \partial_{\tau_1}\varphi_{l}+\beta\partial_\xi\left(\varphi_2\varphi_{l}\right)
+\gamma\partial_\xi^3\varphi_{l}=\Xi_{l}\left(\varphi_2,\ldots,\varphi_{l-1}\right).
\label{kdvl}\end{equation}
Equation (\ref{kdvl}) is the linearized KdV equation obtained by the linearization
of the KdV equation~(\ref{kdv}) about its solution $\varphi_2$ with an additional
right hand member $ \Xi_{l}(\varphi_2,\ldots$, $\varphi_{l-1})$
 depending on the previously determined terms.

This is true with no further hypothesis than the existence of some simple eigenvalue~$V$
of $\Pi_0(A-V)\Pi_0$ and conditions~(\ref{con}) and~(\ref{e11.11}).
However, in the general case, due to the
term $\int^\xi B(u_p,u_{n+1-p})$ in the recurrence formula (\ref{e15.1}),
the right hand member $\Xi_l$ involves {\it a priori} many integrations
 relative to the variable $\xi$. The  property
\begin{equation} Q_1 B(\cdot,\cdot)\equiv0,\end{equation}
satisfied in ferromagnets, ensures their vanishing.

\section{The second order time evolution  and the KdV hierarchy}
\subsection{The problem of the secularities}
It is well known that the KdV equation (\ref{kdv})
 is completely integrable,
i.e.\ that the Cauchy problem for it can be solved by use of the
Inverse Scattering Transform (IST)
method~\cite{Dodd,kdvint,kdvint2}. The IST method gives also
explicit formulas for the resolution of the linearized KdV
equation~(\ref{kdvl}) \cite{kodamah,secular}. This latter equation
is linear, but it is necessary to solve it by this method, because
the solution~$\varphi_2$ of the KdV equation intervenes in it as
an essential parameter. In the general case~$\varphi_2$ can be
only expressed in terms of its inverse transform, thus also the
solution  of the linearized KdV equation~(\ref{kdvl}).
Unfortunately the solution computed this way is not always bounded
(I mean uniformly bounded in~$\xi$ as $\tau\to +\infty$). As an
example, assume that at $\tau=0$, $\varphi_l\equiv 0$, and replace
the right hand member
$\Xi_l\left(\varphi_2,\ldots,\varphi_{l-1}\right)$ by
$\partial_{\xi}\varphi_2$. Then the solution $\varphi_l$ of
equation (\ref{kdvl}) is \cite{kodamah,secular}
\begin{equation} \varphi_l=\tau\partial_\xi \varphi_2.\label{79}\end{equation}
A solution like  (\ref{79}) is called ``secular''. This phenomenon  occurs when
a term in the right hand member ``resonates'', that is, from the mathematical point of view,
 is a solution of the homogeneous equation
 ((\ref{kdvl}) with $\Xi_l\left(\varphi_2,\ldots,\varphi_{l-1}\right)=0$).
Physically, this resonance phenomenon occurs when each component of the
inverse scattering transform of the source term
(some  part of the right hand member $\Xi_l\left(\varphi_2,\varphi_3,\ldots,\varphi_{l-1}\right)$ )
evolves in time in the same way as the corresponding component in the
 transform of the
main term. In order to  remove the secular terms, we  use the method of Kraenkel,
Manna, and Pereira~\cite{manna1,manna3}
that consists in the introduction of additional evolution equations,
which describe the evolution of the main term
relative to the higher order time variables $\tau_2, \tau_3,\ldots$.
 We  check that, according to the  papers cited, these equations
must be those of the so-called KdV hierarchy,
normalized in order to cancel the linear terms in the right hand member
$\Xi_l\left(\varphi_2,\varphi_3,\ldots,\varphi_{l-1}\right)$.
Indeed  references  \cite{kodamah,manna3} state that the secularity producing terms
are the linear ones. We showed in~\cite{secular}
that they are rather the derivatives of the conserved densities
of the KdV equation, and that the procedure of Kraenkel et al.\
in fact removes completely these latter terms from the right hand member.

\subsection{The next order equation}
The standard  expansion used to derive the KdV equation in hydrodynamics
has some parity relative to~$\varepsilon$.
  As an example, Su~\cite{su} uses  series expansions in integer
powers of some perturbative parameter~$\epsilon$, defined
in such a way that $\epsilon=\varepsilon^2$, $\varepsilon$~being the perturbative
parameter of the present paper.
This corresponds
to a vanishing of all odd terms in the present expansion.
The equation (\ref{kdvl}) governing the $\tau_1$ evolution of $\varphi_n$ is obtained
at order $\varepsilon^{n+3}$, which is even when~$n$ is odd.
Due to the homogeneity properties of the expansion,
the right hand member $\Xi_n\left(\varphi_2,\ldots,\varphi_{n-1}\right)$ contains  only
even  order derivatives of $\varphi_2$, which describe losses.
 Physically, if
the initial system (\ref{3.1}) is conservative, the same feature can be expected
for any equation in the expansion.
Therefore the  even  order derivatives of $\varphi_2$ must vanish.
We assume that the present model has the same property in  the following sense:
$\Xi_3\left(\varphi_2\right)=0$. Thus, with a zero initial condition, $\varphi_3\equiv 0$,
and so on: if all $\varphi_p$ with odd $p$ are zero up to
$(n-2)$, $n$ being odd, then $\Xi_{n}\left(\varphi_2,\ldots,\varphi_{n-1}\right)$ is zero
and, with a zero initial condition, $\varphi_{n}$ is also zero.
Thus,   if the corresponding initial conditions are zero,
all the $\Xi_n$ and all the $\varphi_n$ with odd values
of the integer $n$ are zero.
This occurs in the particular case in which the system (\ref{3.1})
 describes wave propagation in a ferromagnetic
medium. This last feature is  proved, using homogeneity properties, in~\cite{kdv2fer}.
Few results can be proved on this point in the general case.
The right hand  member $\Xi_3\left(\varphi_2\right)$
contains a term $K\partial_\xi^4\varphi_2$, where the
coefficient $K$ is given by
\begin{equation} a_0K=\Pi_1(A-V)E^{-1}Q_0(A-V)\left(1-(A-V)^{-1}Q_1(A-V)\right)a_2.
\label{truc1}\end{equation}
In the same way as the expression (\ref{z4}) for $\omega''$ has been derived,
we can  give
an expression for $\omega^{(4)}=\left.{d^4\omega}/{dk^4}\right\vert_{k=0}$.
 After comparison with (\ref{truc1}), it is seen that
$\omega^{(4)}=24i K$.
Observe that  the right hand member
of (\ref{truc1}) contains only real terms. Thus  $K$ is real and $\omega^{(4)}$ is
purely imaginary.
The requirement that the linear term $K\partial_\xi^4\varphi_2$ in
$\Xi_3\left(\varphi_2\right)$ vanish is equivalent to the requirement
that $\omega^{(4)}$ be real, which must be satisfied if the system is conservative.
This generalizes to the higher even order derivatives
and ensures the vanishing of the linear terms, but we
omit the proof because
we must admit the generalization of this feature to the nonlinear terms.
Note that only the linear terms are suspected to be secularity producing. Thus
the vanishing of the linear term is expected to ensure that all terms in
 the expansion are bounded.
Nonetheless, we make the  assumption above, because it is satisfied
in all the examples of which we know, even if we are not able to prove it in the general case.

 With this hypothesis the second nontrivial equation
of the perturbative expansion is the equation (\ref{kdvl}) for $\varphi_4$.
Its right hand member $\Xi_4\left(\varphi_2\right)$ is polynomial
with regard to~$\varphi_2$ and its derivatives, with the
homogeneity of the terms of order $\varepsilon^7$ in the expansion and can thus
be expanded  according to formula
 (\ref{88}) as
\begin{equation}
\Xi_4\left(\varphi_2\right)=\sum_{
\begin{array}{c}\scriptstyle(m_j)_{j\geqslant1},\,k\geqslant0\\
\scriptstyle
\sum_{j\geqslant1}2jm_j+k=5\end{array}}\hspace{-8mm}
\Xi\left((m_j)_{j\geqslant1},k\right)
\prod_{j\geqslant1}\left(\int_{-\infty}^{\xi}\partial_{\tau_j}\right)^{m_j}
\partial_\xi^k\varphi_{2}+
{\cal O}_2\label{142}.\end{equation}
(${\cal O}_2$ designates here an expression in $\varphi_2$ without linear terms.)
The indices $(m_j)_{j\geqslant1}$, $k$ in  this sum can take the  values
\begin{equation}
((m_j)_{j\geqslant1},k)=((0,\ldots),5);\;
((1,0,\ldots),3);\;
((2,0,\ldots),1);\;
((0,1,0,\ldots),1)\end{equation}
only.
Furthermore, the coefficient  $\Xi\left((0,1,0,\ldots),1\right)$ of $\partial_{\tau_2}\varphi_2$ is $-1$.
Thus expression (\ref{142}) can be expanded in the following way:
\begin{gather}
\Xi_4\left(\varphi_2\right)=-\partial_{\tau_2}\varphi_2
+\Xi\left((2,0,\ldots),1\right)\left(\int_{-\infty}^{\xi}\partial_{\tau_1}
\right)^2\partial_\xi\varphi_2\notag\\
\phantom{\Xi_4\left(\varphi_2\right)=}{}+\Xi\left((1,0,\ldots),3\right)\left(\int_{-\infty}^{\xi}\partial_{\tau_1}
\right)\partial_\xi^3\varphi_2
+\Xi\left((0,\ldots),5\right)\partial_\xi^5\varphi_2+{\cal O}_2.
\label{143}\end{gather}
As $\varphi_2$ satisfies the KdV equation (\ref{kdv}), we have
\begin{equation}\int_{-\infty}^{\xi}\partial_{\tau_1}
\varphi_2=-\gamma\partial_\xi^2\varphi_2+{\cal O}_2.\end{equation}
Thus
\begin{equation} \Xi_4\left(\varphi_2\right)=-
\partial_{\tau_2}\varphi_2-\gamma_2\partial_\xi^5\varphi_2+{\cal O}_2,
\label{144}\end{equation}
with
\begin{equation}
 -\gamma_2=\Xi\left((2,0,\ldots),1\right)(-\gamma)^2+\Xi\left((1,0,\ldots),3\right)(-\gamma)
+\Xi\left((0,\ldots),5\right).
\label{145}\end{equation}
The coefficient $\gamma_2$ is explicitly computed in the Appendix (equation (\ref{2k10})).

\subsection{The KdV hierarchy}

The  linear terms  computed above are secularity producing.
Because the solution $\varphi_4$ of the linearized KdV equation must be bounded,
 they must vanish. This imposes some evolution equation
for $\varphi_2$  relative to the second-order time variable $\tau_2$, such that
\begin{equation}
\partial_{\tau_2}\varphi_2=-\gamma_2\partial_\xi^5\varphi_2+{\cal O}_2.\label{146}\end{equation}
The nonlinear terms in equation~(\ref{146})
are not free, but imposed  by the compatibility conditions between
the KdV equation  (\ref{kdv}) and
(\ref{146}): the Schwartz conditions
 $\partial{\tau_1}\partial{\tau_2}\varphi_2=\partial{\tau_2}\partial{\tau_1}\varphi_2$.
Kraenkel, Manna, and Pereira~\cite{manna3} have shown
that the only  equation that has the same homogeneity properties as
$\Xi_4$ and that satisfies this condition is the second equation of
the so-called
KdV hierarchy. The same requirements are found at higher orders.
The KdV hierarchy is the following set of equations~\cite{fnt}:
\begin{equation}\partial_{T_n}v=\partial_X {\cal L}^nv\qquad
\mbox{($n$ integer)},\label{147}\end{equation}
where
\begin{equation}{\cal L}=\frac{-1}4\partial_X ^2-v+\frac12
\int^X dX (\partial_X v).\label{147.1}\end{equation}
For $n=1$ it is the KdV equation,
but with values of the coefficients $\beta=3/2$, $\gamma=1/4$.
The coefficients become equal to those of equation  (\ref{kdv}) by setting
\begin{equation} v=\frac{\beta}{6\gamma}\varphi_2,\qquad X=\xi
\qquad\mbox{and}\qquad T_1=4\gamma\tau_1.
\label{149}\end{equation}
For $n=2$ the equation (\ref{147}) of the hierarchy is
\begin{equation} \partial_{T_2}v=\frac1{16}\partial_X ^5v+\frac54(\partial_X v)\partial_X ^2v
+
\frac58v\partial_X ^3v
+
\frac{15}8v^2\partial_X v.\label{148}\end{equation}
The important feature is the existence of the Hirota $\tau$-function \cite{fnt},
that is a function of all the variables $(X ,T_1,T_2,\ldots)$, related to $v$ by
\begin{equation}
v(X ,T_1,T_2,\ldots)=2\partial_X ^2\ln\tau(X ,T_1,T_2,\ldots)
\end{equation}
(take care to avoid confusion between the Hirota $\tau$-function and
the time variables $\tau_j$).
The existence of $\tau$ insures that a solution $v$ of the whole system
yielded by all equations of the hierarchy exists. As the system admits a solution, the
Schwartz conditions are satisfied at any order,
 i.e.~$\partial_{T_j}\partial_{T_p}v=\partial_{T_p}\partial_{T_j}v$
for any~$j$, $p\geqslant1$.
Furthermore  these conditions
are satisfied formally and identically by the equations themselves and not only
for some particular solution.
The  authors cited have checked by formal computation on the first orders that,
apart from the symmetries of the KdV hierarchy, it is the only compatible system
which has the required homogeneity properties and for which the
first equation is the KdV one. The symmetries of the hierarchy are, after those of the KdV
equation, a~free scaling coefficient for each time variable.
We must identify
\begin{equation}
T_2=-16\gamma_2\tau_2\end{equation} with $\gamma_2$ given by equation (\ref{145})
and impose that $\varphi_2$ satisfy the evolution equation
\begin{equation}
\frac{-1}{16\gamma_2}\partial_{\tau_2}\varphi_2=\partial_\xi{\cal L}^2\varphi_2
\label{150}\end{equation}
with
\begin{equation}{\cal L}=\frac{-1}4\partial_\xi^2-\frac{\beta}{6\gamma}\varphi_2
+\frac{\beta}{12\gamma}\int_{-\infty}^\xi d\xi(\partial_\xi\varphi_2).\label{151}\end{equation}
Then $\Xi_4\left(\varphi_2\right)$  no longer contains any linear term.
 This implies,
 due to the procedure, that it  no longer contains any
 secularity producing term~\cite{secular}.

The linear terms in $\varphi_2$ are removed in the same way at each order.
We impose for each $p\geqslant2$ that
\begin{equation} \frac{-1}{(-4)^p\gamma_p}\partial_{\tau_p}\varphi_2=\partial_\xi{\cal L}^p\varphi_2,
\label{152}\end{equation}
($\cal L$ as above) with $\gamma_p$ defined by $\gamma_1=\gamma$ and
\begin{equation}
\gamma_{p+1}=\sum_{
\begin{array}{c}\scriptstyle(m_j)_{1\leqslant j\leqslant p-1},\,k\geqslant0\\
\scriptstyle
\sum_{j=1}^{p-1}2jm_j+k=2p+3\end{array}}\hspace{-8mm}
\Xi\left((m_j)_{1\leqslant j\leqslant p-1},k\right)
\prod_{j=1}^{p-1}(-\gamma_j)^{m_j}.
\label{153}\end{equation}
For $p=1$ equation (\ref{152}) coincides with equation (\ref{kdv}) with $\beta_1=\beta$.
 We use here the same scheme as in the first case $p=2$.
We have
\begin{equation} {\cal L}^p=\left(\frac{-1}4\partial_\xi^2\right)^p+{\cal O}_1.
\label{154}\end{equation}
Equation (\ref{152}) gives
\begin{equation}\partial_{\tau_p}\varphi_2
=-\gamma_p\partial_\xi^{2p+1}\varphi_2+{\cal O}_2.\label{155}\end{equation}
Then the linear term
 in $\varphi_2$ vanishes from
$\Xi_p\left(\varphi_2,\varphi_4,\ldots,\varphi_{p-2}\right)$.
In this way all secularity producing terms due to $\varphi_2$
vanish at any order.

\section{The third order and the linearized KdV hierarchy}

\subsection{The higher order time evolution of the higher order terms}
The third equation of our perturbation expansion is equation (\ref{kdvl}), written for $l=6$,
with its right hand member $\Xi_6\left(\varphi_2,\varphi_4\right)$ given by formula (\ref{88}).
It is of order $\varepsilon^9$. It involves not only
$\varphi_2$, $\varphi_4$ and their space derivatives
but also $\partial_{\tau_1}\varphi_2$, $\partial_{\tau_2}\varphi_2$,
 $\partial_{\tau_3}\varphi_2$, $\partial_{\tau_1}\varphi_4$
 and $\partial_{\tau_2}\varphi_4$
of orders $\varepsilon^5$, $\varepsilon^7$, $\varepsilon^9$, $\varepsilon^7$ and $\varepsilon^9$ respectively.
$\partial_{\tau_1}\varphi_2$ is replaced by space derivatives
and nonlinear terms using the KdV equation (\ref{kdv}).
We make use of the second and third equations of the hierarchy for $\varphi_2$
so that $\partial_{\tau_2}\varphi_2$ and
$\partial_{\tau_3}\varphi_2$  also vanish, replaced by functions of the
space derivatives.  The
right hand member becomes
\begin{gather}
\Xi_6\left(\varphi_2,\varphi_4\right)=\partial_{\tau_2}\varphi_4
+\Xi\left((2,0,\ldots),1\right)\left(\int_{-\infty}^{\xi}\partial_{\tau_1}
\right)^2\partial_\xi\varphi_4\notag\\
\phantom{\Xi_6\left(\varphi_2,\varphi_4\right)=}
{}+\Xi\left((1,0,\ldots),3\right)\left(\int_{-\infty}^{\xi}\partial_{\tau_1}
\right)\partial_\xi^3\varphi_4
+\Xi\left((0,\ldots),5\right)\partial_\xi^5\varphi_4+{\cal O}_2.
\label{156}\end{gather}
(Here ${\cal O}_2$ represents some expression in $\varphi_2$, $\varphi_4$,
their derivatives and primitives, without linear terms).
The $\tau_1$-evolution of $\varphi_4$ is also known, and described by the
linearized KdV equation (\ref{kdvl}), with the right hand member
$\Xi_4\left(\varphi_2\right)$. Using this relation removes the explicit dependency on
 $\partial_{\tau_1}\varphi_4$, and
the right hand member $\Xi_6\left(\varphi_2,\varphi_4\right)$ reduces  to
\begin{equation}\Xi_6\left(\varphi_2,\varphi_4\right)=\partial_{\tau_2}\varphi_4
+\gamma_2\partial_\xi^5\varphi_4+{\cal O}_2.
\label{158}\end{equation}
Two questions arise at this point:
\begin{itemize}\item How is  the $\tau_2$-dependency of $\varphi_4$ defined?
\item Is the right hand member $\Xi_6\left(\varphi_2,\varphi_4\right)$
 secularity producing due to the terms in
$\varphi_4$, in particular the linear ones?
\end{itemize}
$\varphi_4$
is  the solution  of the linearized KdV
equation (\ref{kdvl}), with the right hand
member~$\Xi_4\left(\varphi_2\right)$ and some given  initial data $\varphi_4(\xi,\tau_1=0)$.
It is expressed as  an integral and linear combination
of the squared Jost functions, defined in the solution
of the KdV equation through the IST method \cite{kodamah,secular}.
We denote by~$\varphi_4^{(1)}$ the solution of the homogeneous
linearized KdV equation with the same initial data
$\varphi_4^{(1)}(\xi,\tau_1=0)\equiv\varphi_4(\xi,\tau_1=0)$,
and by~$\varphi_4^{(2)}$ the solution of the linearized KdV equation with the
right hand member $\Xi_4\left(\varphi_2\right)$ and vanishing initial data
$\varphi_4^{(2)}(\xi,\tau_1=0)\equiv0$, so that $\varphi_4=\varphi_4^{(1)}+\varphi_4^{(2)}$.
It is shown in~\cite{secular} that~$\varphi_4^{(1)}$ is secularity producing
while $\varphi_4^{(2)}$ is not.

Since the $\tau_2$-dependency of the solution $\varphi_2$ of KdV is determined,
the squared Jost functions are also known for  all $(\xi,\tau_1,\tau_2)$, and
so are the spectral components of the right hand member $\Xi_4\left(\varphi_2\right)$.
Thus $\varphi_4^{(2)}$ is  completely determined and its $\tau_2$-dependency is
well defined without additional condition.
Furthermore the evolution equation that makes explicit this dependency is not
needed for the computation of the function
$\varphi_4(\xi,\tau_1,\tau_2)$.
The situation is different with  $\varphi_4^{(1)}$, while the initial
condition is {\it a priori} free.
In the previous subsection we saw that the compatibility conditions fixed
the $\tau_2$-dependency
of~$\varphi_2$, apart from a scaling coefficient for this time variable.
The $\tau_2$-dependency of~$\varphi_4^{(1)}$ is determined below
in the same way, but the scaling coefficients are no longer free.

Consider some solution $v_0$ of the KdV equation, and $v_1$
of the homogeneous linearized KdV equation.
We use here the normalization of formulas (\ref{147})--(\ref{147.1}).
The latter equation is
\begin{equation} \partial_{T_1}v_1+\frac32\partial_X(v_0v_1)+\frac14\partial_X^3v_1=0\label{p+q}.\end{equation}
We assume also that  $v_0$ and $v_1$ are smooth functions of all the variables
$X,T_1,T_2,\ldots$,
and that their dependency relative to  each time variable satisfies
the homogeneity properties of the KdV hierarchy.
We saw above that under these conditions the function~$v_0$,
 solution of the KdV equation, satisfies the complete hierarchy, with some scaling
 constant for each time variable.
Consider some small parameter~$\eta$.
The function
\begin{equation}v=v_0+\eta v_1\end{equation}
satisfies the KdV equation apart from a term of order $\eta^2$. Thus it
satisfies the whole hierarchy (still apart from terms of order~$\eta^2$).
By linearization of the $ n^{\rm th}$ equation  of the hierarchy~(\ref{147}),
 we find that
\begin{equation} \partial_{T_n}v_1=\partial_X {\cal D}_nv_1,\label{159}\end{equation}
with
\begin{equation} {\cal D}_n v_1=\left(d_1{\cal L}^{n-1}+{\cal L}d_1{\cal L}^{n-2}+\cdots
+{\cal L}^{n-1}d_1\right)v_0+{\cal L}^{n}v_1\label{160}\end{equation}
and
\begin{equation} d_1=\frac{d{\cal L}}{dv}(v_1)=-v_1+\frac12\int^X dX (\partial_Xv_1).
\label{161}\end{equation}
This  result is in particular valid for $v_1=\varphi_4^{(1)}$,
 but also for the following orders $v_1=\varphi_p^{(1)}$, with analogous notations,
 for any  even $p\geqslant4$.
While the higher order time evolution of the part
$\varphi_4^{(2)}$ of the term of order $\varepsilon^4$ coming from
the right hand member $\Xi_4\left(\varphi_2\right)$ of the
linearized KdV equation is defined by the main order itself, the
evolution of the part~$\varphi_4^{(1)}$ coming from the initial
data is determined in an analogous way as the main term:
$\varphi_4^{(1)}$~satisfies the linearized KdV hierarchy
(\ref{159}).

Formal identities are related to this feature.
They are found by  the following reasoning.  We
assume as previously that  $v_1$ satisfies the homogeneous
KdV equation. We have  seen that $v_1$ must satisfy equations (\ref{159}).
The existence of $v_1$ as a function of all the variables $X ,T_1,T_2,\ldots$ is not in doubt,
because
it is ensured by the existence of the Hirota $\tau$-func\-tion~\cite{fnt}.
Thus we have
\begin{equation}\partial_{T_n}\partial_{T_j}v_1=\partial_{T_j}\partial_{T_n}v_1\label{163}\end{equation}
for any integers $n$ and $j$.
We denote by
$\left(\partial_{T_n}{\cal D}_j\right)$  the $T_n$-partial derivative of ${\cal D}_j$
which depends on $T_n$ through $v$, and by $\partial_{T_n}{\cal D}_j$  the operator
that applies  successively  ${\cal D}_j$ and~$\partial_{T_n}$. With analogous notation,
$\partial_{X}{\cal D}_j\equiv \left(\partial_{X}{\cal D}_j\right)+{\cal D}_j\partial_X$.
With the use of equation~(\ref{159}) equation~(\ref{163}) becomes
\begin{equation}\left((\partial_{T_n}{\cal D}_j)+{\cal D}_j\partial_X{\cal D}_n\right)v_1=
\left((\partial_{T_j}{\cal D}_n)+{\cal D}_n\partial_X{\cal D}_j\right)v_1.\label{164}\end{equation}
As in the case of the KdV hierarchy itself (nonlinearized), equation (\ref{164})
is valid for any function $v_1$. Thus the following  identity holds formally:
\begin{equation}(\partial_{T_n}{\cal D}_j)+{\cal D}_j\partial_X{\cal D}_n=
(\partial_{T_j}{\cal D}_n)+{\cal D}_n\partial_X{\cal D}_j.\label{165}\end{equation}
This identity can also be written as
\begin{equation} \left[\partial_{T_j}-\partial_X{\cal D}_j\,,\,\partial_{T_n}-\partial_X{\cal D}_n\right]=0,
\label{43.5}\end{equation}
where $\left[M,N\right]=MN-NM$ denotes the commutator of the operators $M$ and $N$,
and  the Schwartz conditions $\left[\partial_{T_j}\,,\,\partial_{T_n}\right]=0$ are assumed.
The commutator in equation (\ref{43.5}) is easily computed and
 vanishes due to the Schwartz condition and to the formal identity~(\ref{165}).

Using the scaling definition for $v_0$ and $T_1, T_2,\ldots$,
we can write down the equations
 of the linearized KdV hierarchy for the functions $\varphi_p^{(1)}$ as
\begin{equation}\partial_{\tau_n}\varphi_p^{(1)}
+(-4)^n\gamma_n\partial_\xi{\cal D}_n\varphi_p^{(1)}
=0,\label{167}\end{equation}
with ${\cal D}_n$ defined by (\ref{160})--(\ref{161}).
In the case that $n=2$ and $p=4$ this equation has the explicit form
\begin{gather}
\partial_{\tau_2}\varphi_4^{(1)}=
-\gamma_2\partial_{\xi}^5\varphi_4^{(1)}
\notag\\
\phantom{\partial_{\tau_2}\varphi_4^{(1)}=}{}
-\frac53\frac{\beta\gamma_2}\gamma
\left[
\varphi_2\partial_{\xi}^3\varphi_4^{(1)}
+2\left(\partial_{\xi}\varphi_2\right)\partial_{\xi}^2\varphi_4^{(1)}
+2\left(\partial_{\xi}^2\varphi_2\right)\partial_{\xi}\varphi_4^{(1)}
+\left(\partial_{\xi}^3\varphi_2\right)\varphi_4^{(1)}
\right]
\notag\\
\phantom{\partial_{\tau_2}\varphi_4^{(1)}=}{}
-\frac56\frac{\beta^2\gamma_2}{\gamma^2}
\left[
\varphi_2^2\partial_{\xi}\varphi_4^{(1)}
+2\varphi_2\left(\partial_{\xi}\varphi_2\right)\varphi_4^{(1)}
\right].\label{yeye}\end{gather}

\subsection{All secular terms vanish}
The function  $\varphi_4^{(1)}$
is secularity producing \cite{secular}. The following term $\varphi_6$
would thus be secular if $\varphi_4^{(1)}$ appears in $\Xi_6\left(\varphi_4,\varphi_2\right)$.
Equation (\ref{yeye}) yields
\begin{equation}\partial_{\tau_2}\varphi_4^{(1)}
=-\gamma_2\partial_\xi^{5}\varphi_4^{(1)}+{\cal O}_2.
\end{equation}
Together with equation (\ref{158}), this allows one to compute
the linear part of the right hand member to obtain
\begin{equation}\Xi_6\left(\varphi_4,\varphi_2\right)=
-\partial_{\tau_2}\varphi_4^{(2)}-\gamma_2\partial_\xi^{5}\varphi_4^{(2)}+{\cal O}_2.\end{equation}
 We admit that the only secularity producing terms due to
 $\varphi_4^{(1)}$  are the linear ones and recall that $\varphi_4^{(2)}$ is
not secularity producing.
Thus,
due to the equation of the linearized KdV hierarchy (\ref{yeye}), all
secularity producing terms vanish from the right
hand member $\Xi_6\left(\varphi_4,\varphi_2\right)$
 and $\varphi_6$ is bounded.

In order to justify the same property at any order,
 we have to compute the linear terms in $\varphi_p$
that appear in the expression of $\Xi_{n}\left(\varphi_2,\ldots,\varphi_{n-2}\right)$.
The computation of all linear terms in these right hand members is detailed
in the Appendix. The general expression of the linear part of $\Xi_n$ is
\begin{gather}
\Xi_{n}\left(\varphi_2,\ldots,\varphi_{n-2}\right)\nonumber\\
\qquad {}=\sum_{
\begin{array}{c}\scriptstyle(m_j)_{j\geqslant1},\,k\geqslant0,\,p\geqslant2\\
\scriptstyle
\sum_{j\geqslant1}2jm_j+k+p=n+3\end{array}}\hspace{-8mm}
\Xi\left((m_j)_{j\geqslant1},k\right)
\prod_{j\geqslant1}\left(\int_{-\infty}^{\xi}\partial_{\tau_j}\right)^{m_j}
\partial_\xi^k\varphi_{p}+
{\cal R}.\label{183}\end{gather}
The remainder $\cal R$ contains
nonlinear terms in $\varphi_p$, as well as terms in $\varphi_j$,
$j\neq p$.
From equation (\ref{167}) of the linearized KdV hierarchy we see that
\begin{equation}
\partial_{\tau_j}\varphi_p^{(1)}=-\gamma_j\partial_\xi^{2j+1}\varphi_p^{(1)}+{\cal O}_2.
\label{184}\end{equation}
Thus
\begin{gather} \Xi_{n}\left(\varphi_2,\ldots,\varphi_{n-2}\right)\nonumber\\
\qquad {}=\sum_{
\begin{array}{c}\scriptstyle(m_j)_{j\geqslant1},\,k\geqslant0\\
\scriptstyle
\sum_{j\geqslant1}2jm_j+k+p=n+3\end{array}}\hspace{-8mm}
\Xi\left((m_j)_{j\geqslant1},k\right)
\prod_{j\geqslant1}\left(-\gamma_j\right)^{m_j}
\partial_\xi^{\left(\sum_{j\geqslant1}2jm_j+k\right)}\varphi_{p}^{(1)}+
{\cal R}'.\label{185}\end{gather}
The remainder ${\cal R}'$ contains nonlinear terms, terms in
$\varphi_j$, $j\neq p$, and in the part~$\varphi_p^{(2)}$ of~$\varphi_p$ that comes
from the right hand member $\Xi_p\left(\varphi_2,\ldots,\varphi_{p-2}\right)$.
These latter terms are not secularity producing.
Using the fact that, in the sum  in~(\ref{185}), $m_j=0$ for
$j>s=\left(n+3-p-1\right)/2$, and expression~(\ref{153})
of $\gamma_j$, we find that
$\Xi_{n}\left(\varphi_2,\ldots,\varphi_{n-2}\right)$ reduces to  ${\cal R}'$. This result
is valid for each~$p$, thus $\Xi_{n}\left(\varphi_2,\ldots,\varphi_{n-2}\right)$
does not contain any secularity producing term and $\varphi_n$ is not secular.

\subsection{The higher order time derivatives}

On the other hand
the  expression of the solution $\varphi_4$ of the linearized KdV equation involves its
spectral transform and an explicit computation of its $\tau_2$-derivative
seems quite impossible. However, it is needed for the explicit computation of the
right hand member $\Xi_6\left(\varphi_2,\varphi_4\right)$.
$\partial_{\tau_2}\varphi_4^{(1)}$ is directly given by equation~(\ref{yeye}). We now seek
an analogous expression of $\partial_{\tau_2}\varphi_4^{(2)}$.
With the use of
the definitions of this section, the linearized KdV equation~(\ref{kdvl}) can be written as
\begin{equation} \left(\partial_{\tau_1}+4\gamma\partial_\xi{\cal D}_1\right)
\varphi_4^{(2)}=\Xi_4\left(\varphi_2\right).\end{equation}
We apply the operator $ \left(\partial_{\tau_2}-16\gamma_2\partial_\xi{\cal D}_2\right)$
to both sides of this equation and make use of identity~(\ref{43.5}), to obtain
 \begin{equation} \left(\partial_{\tau_1}+4\gamma\partial_\xi{\cal D}_1\right)
 \left(\partial_{\tau_2}-16\gamma_2\partial_\xi{\cal D}_2\right)\varphi_4^{(2)}=
 \left(\partial_{\tau_2}-16\gamma_2\partial_\xi{\cal D}_2\right)\Xi_4\left(\varphi_2\right).\end{equation}
$ \left(\partial_{\tau_1}+4\gamma \partial_\xi{\cal D}_1\right) $
is the operator on the left hand side  of the linearized KdV equation that
admits a unique solution for a given
initial data.   $\varphi_4^{(2)}$ is defined in such a way that it vanishes
at $\tau_1=0$ for any $\xi$ and $\tau_2$. Thus its $\xi$ and $\tau_2$ derivatives
also vanish and
\begin{equation} \left.\left(\partial_{\tau_2}-16\gamma_2 \partial_\xi{\cal D}_2\right)
\varphi_4^{(2)}\right\vert_{\tau_1=0}\equiv0.\end{equation}
We use the following notation to denote this solution:
\begin{equation}
\left(\partial_{\tau_2}-16\gamma_2 \partial_\xi{\cal D}_2\right)
\varphi_4^{(2)}=
\left(\partial_{\tau_1}+4\gamma \partial_\xi{\cal D}_1\right)^{-1}
\left[\left(\partial_{\tau_2}-16\gamma_2 \partial_\xi{\cal D}_2\right)
\Xi_4\left(\varphi_2\right)\right].\end{equation}
$\partial_{\tau_2}\Xi_4\left(\varphi_2\right)$
is easily computed while the expression of $\Xi_4$ is explicitly known through
the perturbative expansion and $\partial_{\tau_2}\varphi_2$ is determined
by the second equation of the hierarchy~(\ref{150}).
It is
\begin{equation}\partial_{\tau_2}\Xi_4\left(\varphi_2\right)=
-16\gamma_2\frac{d\,\Xi_4\left(\varphi_2\right)}{d\varphi_2}
\partial_\xi{\cal L}^2\varphi_2,\end{equation}
where ${d\,\Xi_4\left(\varphi_2\right)}/{d\varphi_2}$
is the differential operator obtained by linearization of  $\Xi_4\left(\varphi_2\right)$
about $\varphi_2$\,. $\partial_{\tau_2}\varphi_4$ is thus  given by the  formula
\begin{gather}
\partial_{\tau_2}\varphi_4^{(2)}=16\gamma_2 \left(\partial_\xi{\cal D}_2
\varphi_4^{(2)}-
\left(\partial_{\tau_1}+4\gamma \partial_\xi{\cal D}_1\right)^{-1}
\vphantom{\frac{d\,\Xi_4\left(\varphi_2\right)}{d\varphi_2}}\right.\nonumber\\
\left. \phantom{\partial_{\tau_2}\varphi_4^{(2)}=}{}\times
\left[
\frac{d\,\Xi_4\left(\varphi_2\right)}{d\varphi_2}
\partial_\xi{\cal L}^2\varphi_2
+\partial_\xi{\cal D}_2
\Xi_4\left(\varphi_2\right)\right]\right).\end{gather}

More generally $\varphi_p^{(2)}$ is the solution of the linearized
KdV equation with the right hand member $\Xi_p\left(\varphi_2,\ldots,\varphi_{p-2}\right)$
and zero  initial data and
 \begin{gather} \left(\partial_{\tau_1}+4\gamma\partial_\xi{\cal D}_1\right)
 \left(\partial_{\tau_n}-(-4)^n\gamma_n\partial_\xi{\cal D}_n\right)\varphi_p^{(2)}\nonumber\\
\qquad {}=
 \left(\partial_{\tau_n}-(-4)^n\gamma_n\partial_\xi{\cal D}_n\right)
\Xi_p\left(\varphi_2,\ldots,\varphi_{p-2}\right).\end{gather}
Because, for $n\neq1$, $\left(\partial_{\tau_n}-(-4)^n\gamma_n \partial_\xi{\cal D}_n\right)$
 does not contain
the partial derivative~$\partial_{\tau_1}$,
 $ \left. \left(\partial_{\tau_n}-(-4)^n\gamma_n\partial_\xi{\cal D}_n\right)\varphi_p^{(2)}
\right\vert_{\tau_1=0}\equiv0$.
Using the above notations we obtain the following expression of the
higher order time derivative:
\begin{gather}
\partial_{\tau_n}\varphi_p^{(2)}=
(-4)^n\gamma_n \partial_\xi{\cal D}_n
\varphi_p^{(2)}+\left(\partial_{\tau_1}+4\gamma \partial_\xi{\cal D}_1\right)^{-1}\notag\\
\qquad{}\times
\left[
\sum_{2l=2}^{p-2}
\frac{d\,\Xi_p\left(\varphi_2,\ldots,\varphi_{p-2}\right)}{d\varphi_{2l}}
\partial_{\tau_n}\varphi_{2l}
-(-4)^n\gamma_n\partial_\xi{\cal D}_n
\Xi_p\left(\varphi_2,\ldots,\varphi_{p-2}\right)\right].\label{oror}
\end{gather}
For $p=4$ the time derivatives in the right hand member of equation (\ref{oror})
reduce to $ \partial_{\tau_n}\varphi_{2}=-(-4)^n\gamma_n\partial_\xi{\cal L}^n\varphi_2$,
according to equation (\ref{152}). For larger values of $p$, it involves
$ \partial_{\tau_n}\varphi_{2l}$ with $2l\geqslant4$ that divides into
$ \partial_{\tau_n}\varphi_{2l}=
\partial_{\tau_n}\varphi_{2l}^{(1)}+ \partial_{\tau_n}\varphi_{2l}^{(2)}$,
where $ \partial_{\tau_n}\varphi_{2l}^{(1)}$ is given by the linearized
KdV hierarchy (\ref{167}), and $ \partial_{\tau_n}\varphi_{2l}^{(2)}$
by the same equation (\ref{oror}), in a~recurrent way.

We have proved that the introduction
 of the KdV hierarchy
removes all unbounded
or secular solutions from the perturbative expansion and given the expression
of the higher order time derivatives of all terms in the perturbative expansion.

\section{Conclusion}

We have studied  the higher order terms in the perturbative expansion
that describes KdV solitons in a rather general frame, including
electromagnetic wave propagation in ferromagnetic media.
Using various mathematical techniques,
we have been able to write down the equations satisfied
by the quantities of any order in this expansion.
 In every case it is  the linearized KdV equation,
with some right hand member.
We have summarized the known results about this equation.
Unbounded or secular solutions can be removed
by suppressing linear terms in the right hand member of the equations. This is
done by imposing  that the main term satisfy all equations of the KdV hierarchy.
There exist scaling coefficients for  the higher order time
variables in the hierarchy, which are determined by the requirement that
the linear terms in the right hand member of the linearized KdV equation
vanish. They are computed using explicit recurrence
formulas.
 Furthermore the behaviour of the higher order terms relative to the
higher order time variables was not known. These terms
are divided into two parts: one part comes from the initial data, and
its  higher order time evolution
is described by the linearized KdV hierarchy. The other part comes from the
right hand member of the linearized KdV equation that describes its evolution
relative to the main time scale. It is completely determined, including its
higher order time evolution,  by this latter
 equation.

A quasi-explicit formula for the higher order time derivatives of this second part
of the corrective terms has also been given.
It involves also the expression of the
 linearized KdV hierarchy.
The main result is that none of the higher order  terms
 produces any secularity.
Thus the existence of the expansion is established up to any order.

\subsection*{Acknowledgments}
I express my gratitude to $\rm D^r$ Isabelle Gruais (IRMAR, Universit\'e de Rennes,
France) for  her helpful collaboration, and $\rm P^r$ Guy
M\'etivier for fruitful scientific discussions and as Director of the
 laboratory of Partial Differential Equations of the
 Institut de Recherche Math\'ematique de Rennes (IRMAR),
where I  began the present study. I am also indebted to my former
thesis director
$\rm D^r$ Miguel Manna (Laboratoire de Physique Math\'ematique,
Universit\'e de Montpellier II, France),
whose  works on the KdV hierarchy have considerably influenced the present paper
and who gave me most of the  bibliography.

\appendix

\section{Appendix. Computation of the linear terms\\
in the right hand member of the equations \\
of the perturbative expansion}

The right hand member $\Xi_l\left(\varphi_2,\varphi_3,\ldots,\varphi_{l-1}\right)$
of the linearized KdV equation
 (\ref{kdvl})
is defined by
\begin{equation} a_0\Xi_l\left(\varphi_2,\varphi_3,\ldots,\varphi_{l-1}\right)=\Pi_1{{\cal E}\!  q}_{(l+3)}
+a_0\left(\partial_{\tau_1}\varphi_l+\beta
\partial_{\xi}(\varphi_2\varphi_l)+\gamma\partial_{\xi}^3\varphi_l\right)\label{80}\end{equation}
 for each $l\geqslant3$, with
\begin{equation}
{{\cal E}\!  q}_{(n)}=-Eu_n-(A-V)\partial_\xi u_{n-1}-\sum_{j\geqslant1}\partial_{\tau_j}u_{n-2j-1}
+\sum_{p=2}^{n-2}B(u_p,u_{n-p}),\label{80b}\end{equation}
so that the $ n^{\rm th}$ equation (\ref{pieq})--(\ref{e15.1})
of the perturbative scheme is ${{\cal E}\!  q}_{(n)}=0$
(${{\cal E}\!  q}_{(n)} $ is the difference of the right hand member minus the left hand member).
We  write ${\cal O}_2$ for any polynomial expression in the $\varphi_j$
and their derivatives and primitives that does not contain any linear term. This is
analogous to the usual Landau notation
$O(\varphi_j^2)$, except that we do not assume that the
$\varphi_j$ are small in any way.
Note that the use of the KdV equation satisfied by $\varphi_2$,
and so on could change the linear terms in $\Xi_l\left(\varphi_2,\varphi_3,
\ldots,\varphi_{l-1}\right)$.
Thus in order to define the linear part of $\Xi_l\left(\varphi_2,\varphi_3,\ldots,\varphi_{l-1}\right)$
in a unique way, we assume that no use of these properties has been done.
We made the standard hypothesis that,  for zero initial
conditions at these orders, all $\varphi_j$ with an odd value of $j$ are zero, but
it is more convenient for the present formal computation to forget this feature.
Formal first order terms, vector $u_1$ and  function $\varphi_1$, satisfying
equations (\ref{pieq}) and (\ref{qeq}) are introduced, although it can be shown that they are necessary
zero. This is consistent with the
requirement that no use of the equations has been made: we compute formally the
right hand member
without solving the equations in any way.

We write an {\it a priori}
formula for the linear part of $u_n$:
\begin{equation} u_n=\sum_{\begin{array}{c}\scriptstyle \bar m=\left((m_j)_{j\geqslant1},k,l\right)\\
\scriptstyle
m_j,\,k\geqslant0,\,l\geqslant1,\;
 d(\bar m)=n\end{array}}\tilde u\;(\bar m)
\prod_{j\geqslant1}\left(\int^\xi\partial_{\tau_j}\right)^{m_j}
\partial_\xi^k\varphi_{l}\;
+{\cal O}_2,\label{81}
\end{equation}
where the  components of the vector  coefficients $\tilde u(\bar m)$
have to be determined.
The homogeneity properties of $u_n$ are described by the ``degree'' $d(\bar m)$,
defined by
\begin{equation} d\left((m_j)_{j\geqslant1},k,l\right)
=\left(\sum_{j\geqslant1}2j\,m_j\right)\;+k+l.\end{equation}
 No explicit dependency on $n$ has to be written down because $n=d(\bar m)$.
The substitution of formula (\ref{81}) into the recurrence formulas (\ref{qeq})
and (\ref{e15.1}) for $u_n$
yields a  new  recurrence formula that allows one  to compute these coefficients.
 This  shows
by induction that formula (\ref{81}) is valid for any value of the integer~$n$
and that the values found for
 the coefficients  are valid. These recurrence formulas are
\begin{gather} \mbox{For all $l\geqslant1$:}\hspace{3.1cm} \tilde u\;((0),0,l)=a_0
\label{82}\\
\mbox{For all $k$ and $l\geqslant1$:}\hspace{2cm}
 \tilde u\;((0),k,l)=F_1(\tilde u((0),k-1,l)).
\label{83}\end{gather}
 For all $(m_j)_{j\geqslant1}\neq(0)$
and $l\geqslant1$,
\begin{equation}
 \tilde u\;((m_j)_{j\geqslant1},0,l)=
\sum_{i\geqslant1}F_3(\tilde u\;((m_j-\delta_{i,j})_{j\geqslant1},0,l)).
\label{84}\end{equation}
For all $(m_j)_{j\geqslant1}\neq(0)$, $k$, $l\geqslant0$,
\begin{gather}
 \tilde u\;((m_j)_{j\geqslant1},k,l)=
F_1(\tilde u((m_j)_{j\geqslant1},k-1,l))\notag\\
\qquad{}+\sum_{i\geqslant1}
\left[F_2(\tilde u\;((m_j-\delta_{i,j})_{j\geqslant1},k-1,l))
+ F_3(\tilde u((m_j-\delta_{i,j})_{j\geqslant1},k,l))\right].
\label{85}\end{gather}
$\delta_{i,j}$ is the Kronecker symbol  and the operators $F_j$ are defined by
\begin{gather} F_1=F_2(A-V),\\
 F_2=-\left[1-(A-V)^{-1}Q_1(A-V)\right]E^{-1}Q_0,\\
  F_3=-(A-V)^{-1}Q_1.\end{gather}
Formulas (\ref{82}) to (\ref{85}) define
 $\tilde u\;((m_j)_{j\geqslant1},k,l))$.
We see that this quantity  does not depend upon $l$.  We write
more simply $\tilde u\;((m_j)_{j\geqslant1},k))$ in the following.
For some particular terms explicit expressions can be given.
It is straightforwardly seen from equations (\ref{83}) and~(\ref{82}) that, for all $k$,
\begin{equation}\tilde u((0),k)={F_1}^ka_0.\end{equation}
In the particular case that $k=1$ this is
\begin{equation} \tilde u\;((0),1)) =a_1+a'_1.\label{86}\end{equation}
Equations (\ref{84}) and (\ref{82}) yield in a similar way
\begin{equation}\tilde u((m_j)_{j\geqslant1},0)
={F_3}^{\sum_{j\geqslant1}m_j}a_0.\end{equation}
The definition of $a_0$ shows that $Q_1a_0=0$. Thus $F_3a_0$ is
zero and $\tilde u((m_j)_{j\geqslant1},0)=0$ for all $(m_j)_{j\geqslant1}$.
In the same way, due to $Q_0a_0=0$, $F_2a_0=0$.

 The  formulas obtained are substituted into the definition (\ref{80b}) of
$\Xi_l$ to give
\begin{gather}
\Pi_1{{\cal E}\!  q}_{(n)}=
-\sum_{
\begin{array}{c}
\scriptstyle  (m_j)_{j\geqslant1},\,k\,\geqslant0\,,l\,\geqslant1\\
\scriptstyle d(\bar m)=n-1
\end{array}}
\Pi_1(A-V)Q_0\tilde u(\bar m)
\prod_{j\geqslant1}\left(\int_{-\infty}^{\xi}\partial_{\tau_j}\right)^{m_j}
\partial_\xi^{k+1}\varphi_{l}\notag
\\
\phantom{\Pi_1{{\cal E}\!  q}_{(n)}=}{}-\sum_{i\geqslant1}\!
\sum_{
\begin{array}{c}\scriptstyle(m_j)_{j\geqslant1},\,k\,\geqslant0\,,l\,\geqslant1\\
\scriptstyle d(\bar m)=n-2i-1
\end{array}}\!\!
\Pi_1\tilde u(\bar m)
\prod_{j\geqslant1}
\left(\int_{-\infty}^{\xi}\partial_{\tau_j}\right)^{m_j+\delta_{i,j}}
\partial_\xi^{k+1}\varphi_{l}+
{\cal O}_2.
\label{87}\end{gather}
Thus the right hand member $\Xi_n$ is
\begin{gather}\Xi_n\left(\varphi_2,\varphi_3,\ldots\varphi_{n-1}\right)\nonumber\\
\qquad {}=\sum_{
\begin{array}{c}\scriptstyle(m_j)_{j\geqslant1},\,k\geqslant0\\
\scriptstyle1\leqslant l\,\leqslant n-1\\
\scriptstyle
\sum_{j\geqslant1}2jm_j+k+l=n+3
\end{array}}\hspace{-8mm}
\Xi\left((m_j)_{j\geqslant1},k\right)
\prod_{j\geqslant1}\left(\int_{-\infty}^{\xi}\partial_{\tau_j}\right)^{m_j}
\partial_\xi^k\varphi_{l}+
{\cal O}_2,\label{88}\end{gather}
with
\begin{gather} a_0\Xi\left((m_j)_{j\geqslant1},k\right)
=-\Pi_1(A-V)Q_0\tilde u\left((m_j)_{j\geqslant1},k-1\right)\nonumber\\
\phantom{a_0\Xi\left((m_j)_{j\geqslant1},k\right)=}{}
 -\sum_{i\geqslant1}\Pi_1
\tilde u\left((m_j-\delta_{i,j})_{j\geqslant1},k-1\right),\label{89}\end{gather}
where $\tilde u((m_j)_{j\geqslant1},k)$ is defined by
 the recurrence
formulas (\ref{82}) to (\ref{85}).
Note the remarkable feature that the coefficient $\Xi\left((m_j)_{j\geqslant1},k\right)$
does not depend on $n$.

The term in which $\partial_{\tau_j}\varphi_l$ appears with
the largest
value of the index $j$, in a given~$\Xi_n$, can be computed explicitly.
 Because $n$
is necessary even, we write  $n=2p$. Due to the homogeneity
properties of $\Xi_n\left(\varphi_1,\ldots\right)$
 we see that  the term sought is proportional to~$\partial_{\tau_p}\varphi_2$. The
corresponding coefficient is $\Xi \left((\delta_{j,p}),1\right)$.
By direct application of the previous formulas, we find that
\begin{equation}\Xi\left((\delta_{j,p}),1\right)=-1.\label{90}\end{equation}
The coefficient $\gamma_2$ that defines the second order time
scale $\tau_2$ is computed using the expansion (\ref{145}) and the
above recurrence formulas. It is
\begin{equation}\gamma_2a_0=\Pi_1\left[-\gamma F_1+(A-V)Q_0\left(
{F_1}^3-\gamma\left[ F_1F_3+F_3F_1+F_2\right]\right)\right]F_1a_0\label{2k10}.\end{equation}

\label{Leblond-lastpage}

\end{document}